\documentclass[12pt, a4paper]{article}
\usepackage[english]{babel}
\usepackage[utf8x]{inputenc}
\usepackage{amsmath}
\usepackage{graphicx}
\usepackage{lmodern}
\usepackage{textcomp}
\usepackage[colorinlistoftodos]{todonotes}

\usepackage[mathscr]{euscript}
\newcommand\independent{\protect\mathpalette{\protect\independenT}{\perp}}
\def\independenT#1#2{\mathrel{\rlap{$#1#2$}\mkern2mu{#1#2}}}

\usepackage{bm}
\usepackage{amsthm}

\theoremstyle{definition}
\newtheorem{definition}{Definition}[]

\usepackage{booktabs,array}

\usepackage[plain,noend]{algorithm2e}

\usepackage{amsmath}
\usepackage{graphicx}

\usepackage{amssymb}

\usepackage[mathscr]{euscript}

\usepackage{titlesec}
\titleformat{\section}[block]
  {\center}
  {\thesection .}
  {1em}
  {\MakeUppercase}
\titleformat{\subsection}[hang]
  {\center \it}
  {\thesubsection}
  {1em}
  {}
\usepackage{authblk}
\author[1]{Anders Huitfeldt}
\author[2]{Mats J. Stensrud}
\author[3]{Etsuji Suzuki}
\affil[1]{The Meta-Research Innovation Center at Stanford, Stanford University}
\affil[2]{Department of Biostatistics, University of Oslo}
\affil[3]{Department of Epidemiology, Okayama University}

\begin{document}

\title{On the collapsibility of measures of effect in the counterfactual causal framework}

\maketitle

\begin{abstract}

The relationship between collapsibility and confounding has been subject to an extensive and ongoing discussion in the methodological literature.  We discuss two subtly different definitions of collapsibility, and show that by considering causal measures of effect based on counterfactual variables it is possible to separate out the component of non-collapsibility which is due to the mathematical properties of the effect measure, from the components that are due to structural bias such as confounding. We provide weights such that the causal risk difference and the causal risk ratio are collapsible over arbitrary baseline covariates, and demonstrate that such general weights do not exist for the odds ratio. 
%
%\subsection*{Conclusions}
%
%The relationship between confounding and non-collapsiblity depends on the definition of collapsibility that is being considered. Using a definition based on counterfactual effect measures allows the construction of simple, previously unpublished weights which may be used for standardization of causal effects.  

\end{abstract}

\section{Introduction}

A measure of association (such as the risk difference or the risk ratio) is said to be collapsible if the marginal measure of association is equal to a weighted average of the stratum-specific measures of association \cite{Whittemore1978CollapsibilityTables}. The relationship between collapsibility and confounding has been subject to an extensive and ongoing discussion in the literature\cite{Greenland1999ConfoundingInference}. In this paper, we argue that the concept of collapsibility can be made clearer by framing the discussion in terms of causal effect measures based on counterfactual variables. 

In all the examples, we are interested in the effect of a binary exposure \(A\) (e.g. a drug), on a binary outcome \(Y\) (e.g. a side effect). We use superscript to denote counterfactual variables \cite{Hernan2016CausalInference}. For example, \(Y^{a=1}\) is an indicator for whether an individual would have got the outcome if, possibly contrary to fact, she had been exposed to the drug. We will make a distinction between measures of association, which compare the distribution of the outcomes in the exposed with the distribution of outcomes in the unexposed; and causal measures of effect, which compare the counterfactual distribution under exposure (across everyone) with the counterfactual distribution under the absence of exposure (across everyone). For example, the associational risk difference is \( \text{Pr}(Y=1 \vert A=1) - \text{Pr}(Y=1\vert A=0)\) whereas the causal risk difference ($RD$) is \(\text{Pr}(Y^{a=1}=1)-\text{Pr}(Y^{a=0}=1)\). These effect measures may be defined within levels of covariates \(V\). We denote this using a subscript on the effect measure: $RD_v = \text{Pr}(Y^{a=1}=1 \vert V=v)-\text{Pr}(Y^{a=0}=1\vert V=v)$.

\section{Definitions of Collapsibility}

We will adopt Pearl's definition of collapsibility for measures of association \cite{Pearl2009Causality:Inference}:

\begin{definition}{(Collapsibility of a Measure of Association)}
Let \(g[f(A,Y)] \) be any functional that measures the association between \(A\) and \(Y\) in the joint distribution \(f(A,Y)\). We say that \(g\) is collapsible on a variable \(V\) with weights \(w_v\) if \(\frac{\sum_v{\{g[f(A,Y \vert V=v)]\times w_v}\}} {\sum_v{w_v}} = g[f(A,Y)] \)
\end{definition}

Newman \cite{Newman2001BiostatisticalEpidemiology} showed  conditions under which the associational risk difference, risk ratio, and odds ratio are collapsible according to this definition. He also provided corresponding weights. Briefly, we note that: the associational risk difference is collapsible with weights \(\text{Pr}(V=v \vert A=1)\) if \(V\) is not associated with the outcome in the unexposed, or if \(V\) is not associated with the exposure; the associational risk ratio is collapsible with weights \(\text{Pr}(V=v \vert A=1)\times \text{Pr}(Y=1 \vert A=0, V=v)\) under similar conditions; and the associational odds ratio is collapsible with weights \( \text{Pr}(V=v \vert A=1)\times \text{Pr}(Y=0 \vert A=1,V=v)\times \frac{\text{Pr}(Y=1 \vert A=0, V=v)}{1-\text{Pr}(Y=1 \vert A=0,V=v)} \) under certain very limiting conditions, for example if \(\text{Pr}(Y=1\vert A=0, V=v)\) is equal for all values of \(v\). A full discussion of the graphical and probabilistic conditions that lead to collapsibility under this definition is provided by Greenland and Pearl \cite{Greenland2011AdjustmentsModels}.

From these results, it follows that general statements about the collapsibility properties of effect measures (e.g. ``the risk difference is collapsible'') must either be qualified by the specification of the conditions that are being assumed, or alternatively taken to refer to some other definition of collapsibility. We propose a suitable definition:  a causal measure of effect is collapsible if the marginal effect measure is equal to a weighted average of the stratum-specific causal effect measures. This is a formalization of the definition used in Fine Point 4.3 in Hernan and Robins textbook \textit{Causal Inference}:

\begin{definition}{(Collapsibility of a Measure of Causal Effect)}

Let \(h[f(Y^{a=0},Y^{a=1})]\) be any functional that measures the association between \(Y^{a=0}\) and \(Y^{a=1}\) in the joint distribution \(f(Y^{a=0}, Y^{a=1})\). We say that \(h\) is collapsible on a variable \(V\) with weights \(w_v\) if \(\frac{\sum_v{\{\{h[f(Y^{a=0},Y^{a=1} \vert V=v)]\times w_v}\}} {\sum_v{w_v}} = h[f(Y^{a=0},Y^{a=1})] \)
\end{definition}

Under definition 2, collapsibility is understood as a mathematical property of the effect measure, rather than a consequence of certain graphical or probabilistic structures in the data set. Consequently, results from Greenland and Pearl do not apply under definition 2, and measures of effect may be collapsible over \(V\) even if \(V\) is a confounder. Definitions 1 and 2 are not generally equivalent: a set of weights that satisfies definition 1 may not satisfy definition 2, and conversely a set of weights that satisfies definition 2 may not satisfy definition 1. The definitions are however equivalent if there is both no confounding conditional on \textit{V}, and no confounding unconditionally  (i.e. if \(Y^{a} \independent A\) and \(Y^{a} \independent A \vert V\) for all values of \textit{a}). 

Finally, we consider a third related concept, discussed by Miettinen \cite{Miettinen1972StandardizationRatios.}, who stated (correctly, but without proof) that the ``standardized risk ratio'' (SRR), which is constructed by standardizing the risk in the exposed and the risk in the unexposed separately with weights Pr(\(V=v\)) and reporting the ratio of these measures (Formula 4 in Miettinen), is equal to a weighted average of the stratum-specific risk ratios under the weights \( \text{Pr}(V=v)\times \text{Pr}(Y=1 \vert A=0, V=v).\) (Formula 6 in Miettinen). Since Miettinen’s SRR is equal to the causal risk ratio if there is no unmeasured confounding, Miettinen’s weights satisfy Definition 2 in the special case of no confounding conditional on $V$.

\section{Collapsibility of Measures of Causal Effect}

\subsection{Risk difference}

The causal risk difference is collapsible over covariates \textit{V} with respect to weights \(w_v\) if \(\frac {\sum_v{\{[\text{Pr}(Y^{a=1}=1|V=v)-\text{Pr}(Y^{a=0}=1|V=v)]\times w_v}\}}{\sum_v{w_v}} = \text{Pr}(Y^{a=1}=1)-\text{Pr}(Y^{a=0}=1) \). We next proceed to show that the causal risk difference is collapsible over arbitrary covariates $V$ if we use the weights $w_v = \text{Pr}(V=v)$.

First note that the sum of the weights is 1, allowing the denominator to be ignored. Next,

\begin{equation}
\begin{aligned}
&\sum_v{[\text{Pr}(Y^{a=1}=1|V=v)-\text{Pr}(Y^{a=0}=1|V=v)]\times \text{Pr}(V=v)}  \\
&=  \sum_v{\text{Pr}(Y^{a=1}=1|V=v)\times \text{Pr}(V=v)}- \sum_v{\text{Pr}(Y^{a=0}=1|V=v)\times \text{Pr}(V=v)} \\
&=  \text{Pr}(Y^{a=1}=1)-\text{Pr}(Y^{a=0}=1) 
\end{aligned}
\end{equation}

Also note that if the risk difference is the same in every stratum (i.e. in the absence of effect modification) the stratum-specific risk differences will also be equal to the marginal risk difference, and the risk difference is collapsible with any weights. It can be shown that this is true for any measure of effect for which there exist weights that guarantee collapsibility over arbitrary covariates. 

\subsection{Risk Ratio}  

The risk ratio is asymmetric with respect to coding of the outcome, so it is necessary to consider each risk ratio model separately. These are defined as follows: 
\bigskip

\begin{center}
\(  RR(-) = \frac{\text{Pr}(Y^{a=1}=1)}{\text{Pr}(Y^{a=0}=1)}      \)    
\\
\bigskip
\( RR(+) = \frac{\text{Pr}(Y^{a=1}=0)}{\text{Pr}(Y^{a=0}=0)} \)
\end{center}
\bigskip

The two risk ratio models require different sets of weights for collapsibility. We next show that the causal risk ratio \(RR(-)\) is collapsible over arbitrary covariates $V$ if we use the weights $w_v = \text{Pr}(V=v \vert Y^{a=0}=1)$, i.e. weights determined by the distribution of the baseline covariates among those individuals who would have been cases if they, possibly contrary to fact, were not treated with drug \textit{A}:
\bigskip

Our goal is to show that \[\frac{\text{Pr}(Y^{a=1}=1)}{\text{Pr}(Y^{a=0}=1)} = \frac
{\sum_v{[\frac{\text{Pr}(Y^{a=1}=1|V=v)}{\text{Pr}(Y^{a=0}=1|V=v)}\times \text{Pr}(V=v \vert (Y^{a=0}=1)]}}{\sum_v{\text{Pr}(V=v \vert Y^{a=0}=1)}}\]

Again, we note that the sum of the weights is 1, and that the denominator can therefore be ignored. 

\begin{equation}
\begin{aligned}
&\sum_v{\frac{\text{Pr}(Y^{a=1}=1|V=v)\times \text{Pr}(V=v \vert Y^{a=0}=1)}{\text{Pr}(Y^{a=0}=1\vert V=v)}}\\
&=
\sum_v{\frac{\text{Pr}(Y^{a=1}=1|V=v)\times \text{Pr}(Y^{a=0}=1|V=v)\times \text{Pr}(V=v)}{\text{Pr}(Y^{a=0}=1|V=v)\times \text{Pr}(Y^{a=0}=1)}}&\text{(Bayes Theorem)}\\
&=
\sum_v{\frac{\text{Pr}(Y^{a=1}=1|V=v)\times \text{Pr}(V=v)}{\text{Pr}(Y^{a=0}=1)}}\\
&=
\frac{\sum_v{\text{Pr}(Y^{a=1}=1|V=v)\times \text{Pr}(V=v)}}{\text{Pr}(Y^{a=0}=1)} &\\
&=
\frac{\text{Pr}(Y^{a=1}=1)}{\text{Pr}(Y^{a=0}=1)} &
\end{aligned}
\end{equation}

This proof is not invariant to the coding of the exposure or outcome variables, and the correct weights will therefore depend on the exact specification of the risk ratio parameter.  Analogous proofs can be provided to show that the weights for \(RR(+)\) are given by $\text{Pr}(V=v \vert Y^{a=0}=0)$, the weights for $\frac{1}{RR(-)}$ are given by $\text{Pr}(V=v \vert Y^{a=1}=1)$, and that the weights for $\frac{1}{RR(+)}$ are given by $\text{Pr}(V=v \vert Y^{a=1}=0)$, 

Note that the marginal causal risk ratio is generally not equal to a weighted average of the conditional causal risk ratios, if the weights are determined by the marginal distribution of the covariates $V$. Exceptions occur in special situations, such as when the risk ratio is equal in every stratum (i.e. when there is no effect modification on the risk ratio scale).

\subsection{Odds Ratio}

For all the previously discussed parameters, we have shown that for any baseline covariates \textit{V}, there exist weights such that the marginal effect measure is equal to a weighted average of the stratum-specific effects. We will now show that this does not hold for the odds ratio by considering the following simple counterexample:

Consider a population, with 25 percent men and 75 percent women, where a randomized trial is conducted on the effect of drug \textit{A}. The hypothetical results are shown in Table 1. The randomization probability is equal in men and women and we have an infinite sample size, there is therefore no confounding. 

\begin{table}
\caption{Conditional and Marginal Odds Ratios}
\begin{tabular}{p{0.3\linewidth}p{0.2\linewidth}p{0.2\linewidth}p{0.2\linewidth}}
\\
&  \textit{Average Counterfactual Risk of Outcome (Placebo)} &  \textit{Average Counterfactual Risk of Outcome (Treatment)} &  \textit{Odds Ratio}\\[5pt]
Men (25 Percent) & 0.5 & 0.75 & 3\\
Women (75 Percent) & 0.25 & 0.5 & 3 \\
Overall & 0.3125 & 0.5625 & 2.82 \\
\end{tabular}
\label{tablelabel}
\end{table}

This table shows that for the variable sex, the stratum-specific causal odds ratios are equal between men and women, but the overall causal odds ratio is different from the stratum-specific odds ratios. Moreover, since any weighted average of the stratum-specific odds ratios is 3, there does not exist any set of weights that makes the odds ratio collapsible over sex. This counterexample shows that no generally applicable weights such as those for the risk difference and the risk ratio can be provided for the odds ratio.  
 
\section{Identification of the Weights}

If the investigator intends to report an average of the stratum-specific effects as an estimate of the marginal effect, it is necessary to know not only that the effect is collapsible in principle, but also to construct appropriate weights, identify them from the data and apply them in the analysis. The weights for the risk ratio $RR(-)$, Pr$(V=v \vert Y^{a=0}=1)$, have a counterfactual variable in the conditioning event, and may not be identified from the data. However, we proceed to show that the weights are identified in the absence of unmeasured confounding, \textit{i.e} if $Y^{a=0} \independent A \vert V$
\bigskip

\begin{proof} 

\begin{equation}
\begin{aligned}
&\text{Pr}(V=v|Y^{a=0}=1)  \\
&= \frac{\text{Pr}(Y^{a=0}=1 \vert V=v)\times \text{Pr}(V=v)}{\text{Pr}(Y^{a=0}=1)} &\text{Bayes Theorem}\\
&= \frac{\text{Pr}(Y^{a=0}=1 \vert A=0, V=v)\times \text{Pr}(V=v)}{\text{Pr}(Y^{a=0}=1)} &\text{Exchangeability}\\
&= \frac{\text{Pr}(Y=1 \vert A=0, V=v)\times \text{Pr}(V=v)}{\text{Pr}(Y^{a=0}=1)} &\text{Consistency}\\
\end{aligned}
\end{equation}
\end{proof}

\(\text{Pr}(Y^{a=0}=1)\) is constant over $v$ and can therefore be factored out of the weights. In the absence of confounding, the weights Pr$(V=v \vert Y^{a=0}=1)$ are therefore equivalent to Miettinen’s weights Pr$(V=v)\times \text{Pr}(Y=1 \vert A=0, V=v)$ as discussed earlier.

An alternative identification of the weights can be used if standardizing experimental results to a population where everyone is unexposed. In such situations, \(Y^{a=0}= Y\) in all individuals by consistency, and the weights in the target population are identified as \(\text{Pr}(V=v \vert Y=1)\)

\section{Discussion}

We have reviewed well-established results from previous work on the collapsibility of measures of association, and shown corresponding results for causal measures of effect. With these causal effect measures, one is able to disentangle the components of non-collapsibility that are due to the mathematical properties of the effect measure from the components that are due to structural bias and the probabilistic structure of the dataset. We have provided new, simple weights for the causal risk ratio, which guarantee collapsibility over arbitrary baseline covariates, and showed that such weights do not exist for the causal odds ratio. 

Our weights for the risk ratio $RR(-)$ are equivalent to the weights previously discussed by Miettinen when there is no unmeasured confounding; in other words, in all situations where standardizing over \textit{V} provides a valid estimate of the causal effect. However, our formulation allows much simpler presentation of the weights and of the proofs. Furthermore, our formulation highlights pitfalls of using weighted averages: When conditioning on \textit{V}, the correct weights cannot be estimated from the data if unmeasured confounding is present. In such scenarios, using erroneous weights may amplify the bias that is caused by unmeasured confounding within the strata. 

Finally, we note that in many cases it is possible to sidestep the collapsibility of the effect measure entirely, by standardizing the distributions of $Y^{a=1}$ and $Y^{a=0}$ separately. One way to do this is by reporting the overall marginal risk ratio $RR(-)$ as \[ \frac{\sum_v \text{Pr}(Y^{a=1}=1\vert V=v)\times \text{Pr}(V=v)}{\sum_v \text{Pr}(Y^{a=0}=1\vert V=v)\times \text{Pr}(V=v)}\]
Since this procedure does not depend on non-collapsibility, analogous procedures are valid for any effect measure, including the odds ratio.

\bibliographystyle{unsrt}
\bibliography{Mendeley.bib}

\section*{Acknowledgement}
The authors thank James Robins for pointing out the link between the weights proposed in this paper, and previously published weights due to Miettinen. 

\section*{Author Contributions} AH had the original idea, provided the original version of the theorems and proofs, wrote the first draft of the manuscript and coordinated the research project. MJS and ES contributed original intellectual content and extensively restructured and revised the manuscript. All authors approved the final version of the manuscript.

\end{document}